\documentclass[showpacs,superscriptaddress,amsmath,amssymb,nofootinbib,twocolumn,floatfix,reprint]{revtex4-1}

\usepackage{multirow}

\usepackage{setspace}
\usepackage[hidelinks]{hyperref}
\usepackage{graphicx}
\newcommand{\msun}{\rm {M\textsubscript{\(\odot\)}}}

\newcommand{\dlumi}{D_{\rm{L}}}

\newcommand{\rr}{\mathcal{R}_{\rm{90}}}
\newcommand{\td}{{\rm d}}

\begin{document}

%\begin{center}
% LIGO-P2200184-v2
%\end{center}

% ======================
% TITLE AND ABSTRACT
% ======================

\title{Searches for Mass-Asymmetric Compact Binary Coalescence Events using Neural Networks in the LIGO/Virgo Third Observation Period}

%\author{M. Andr\'es-Carcasona\,\orcidlink{0000-0002-8738-1672}}
%\email{mandres@ifae.es (corresponding author)}
%\affiliation{Institut de Física d’Altes Energies (IFAE), Barcelona Institute of Science and Technology, E-08193 Barcelona, Spain}
%\author{A. Men\'endez-V\'azquez\,\orcidlink{0000-0002-0828-8219}}
%\affiliation{Institut de Física d’Altes Energies (IFAE), Barcelona Institute of Science and Technology, E-08193 Barcelona, Spain}
%\author{M. Mart\'inez\,\orcidlink{0000-0002-3135-945X}}
%\affiliation{Institut de Física d’Altes Energies (IFAE), Barcelona Institute of Science and Technology, E-08193 Barcelona, Spain}
%\affiliation{Catalan Institution for Research and Advanced Studies (ICREA), E-08010 Barcelona, Spain}
%\author{Ll. M. Mir\,\orcidlink{0000-0002-4276-715X}}
%\affiliation{Institut de Física d’Altes Energies (IFAE), Barcelona Institute of Science and Technology, E-08193 Barcelona, Spain}

\author{M. Andr\'es-Carcasona}
%\email{mandres@ifae.es (corresponding author)}
\affiliation{Institut de Física d’Altes Energies (IFAE), Barcelona Institute of Science and Technology, E-08193 Barcelona, Spain}
\author{A. Men\'endez-V\'azquez}
\affiliation{Institut de Física d’Altes Energies (IFAE), Barcelona Institute of Science and Technology, E-08193 Barcelona, Spain}
\author{M. Mart\'inez}
\affiliation{Institut de Física d’Altes Energies (IFAE), Barcelona Institute of Science and Technology, E-08193 Barcelona, Spain}
\affiliation{Catalan Institution for Research and Advanced Studies (ICREA), E-08010 Barcelona, Spain}
\author{Ll. M. Mir}
\affiliation{Institut de Física d’Altes Energies (IFAE), Barcelona Institute of Science and Technology, E-08193 Barcelona, Spain}

\date{\today}

% ----- abstract

\begin{abstract}
We present the results on the search for the coalescence of compact binary mergers with very asymmetric mass configurations 
using convolutional neural networks  and 
the  LIGO/Virgo data for the O3 observation period. Two-dimensional images in time and frequency are used as input.  
Masses in the range between 0.01  $\msun$ and 20 $\msun$
are considered. 
We explore neural networks trained with input information from a single interferometer, pairs of interferometers, or all three 
interferometers together,  indicating that the use of the maximum information available  
leads to an improved performance. A scan over the O3 data set using the 
convolutional neural networks for detection results into no significant excess from an only-noise hypothesis. 
The results are translated into  90$\%$ confidence level upper limits on the merger rate as a function of the mass parameters of the binary system. 
\end{abstract}

\pacs{95.85.Sz, 04.80.Nn, 95.55.Ym, 04.30-w, 04.30.Tv}  % PACS, the Physics and Astronomy

\maketitle

% ========================
% INTRODUCTION
% ========================
\section{Introduction}
\label{sec:intro}

Since the discovery of Gravitational Waves (GW) in 2015~\cite{FirstGWDet}, generated by a compact 
binary coalescence (CBC) of black holes (BH), the LIGO and Virgo experiments have improved their sensitivity and observed an increasing number of GW signals, including also  events attributed to the coalescence of neutron stars (NS), as well as the coalescence of BH-NS binary systems.  The latest catalogue of events, from O1, O2 and O3 observation runs, collects a total of 90 events, dominated by BH-BH candidates~\cite{Abbott2019GWTC-1:Runs,Abbott2020GWTC-2:Run,Abbott2021GWTC-3:Run}. The data indicate that the masses in the binary systems range between $1.17$\ \msun (GW191219\_163120) and $105$ \msun (GW190426\_190642), with a mass ratio $q \equiv m_1/m_2$, where $m_1$ denotes the heaviest of the two objects, in the range between $1.1$ (GW170817) and $26.5$ (GW191219\_163120). The LIGO and Virgo Collaborations use matched-filtering techniques to extract the events from the much larger background (for a comprehensive review of the experimental techniques see Ref.~\cite{LIGOScientific:2019hgc}). The use of machine learning tools has been extensively explored in LIGO and Virgo (for a comprehensive review see Ref.~\cite{Cuoco_2020}). In particular, the presence of a distinct chirp-like shape in the CBC events, 
when represented in spectrograms showing the signal in  frequency-time domain, 
makes the use of a convolutional neural network (CNN)  a valid alternative suitable for GW detection~\cite{PhysRevLett.120.141103,george2017deep,Gebhard_2019,George2018DeepData,PhysRevD.103.062004}.  
In addition, the use of CNNs has been explored to distinguish between families of glitches~\cite{Razzano_2018,Biswas_2013,Cavaglia_2019}.

\begin{figure}[htbp]
    \centering
    \includegraphics[width=\columnwidth]{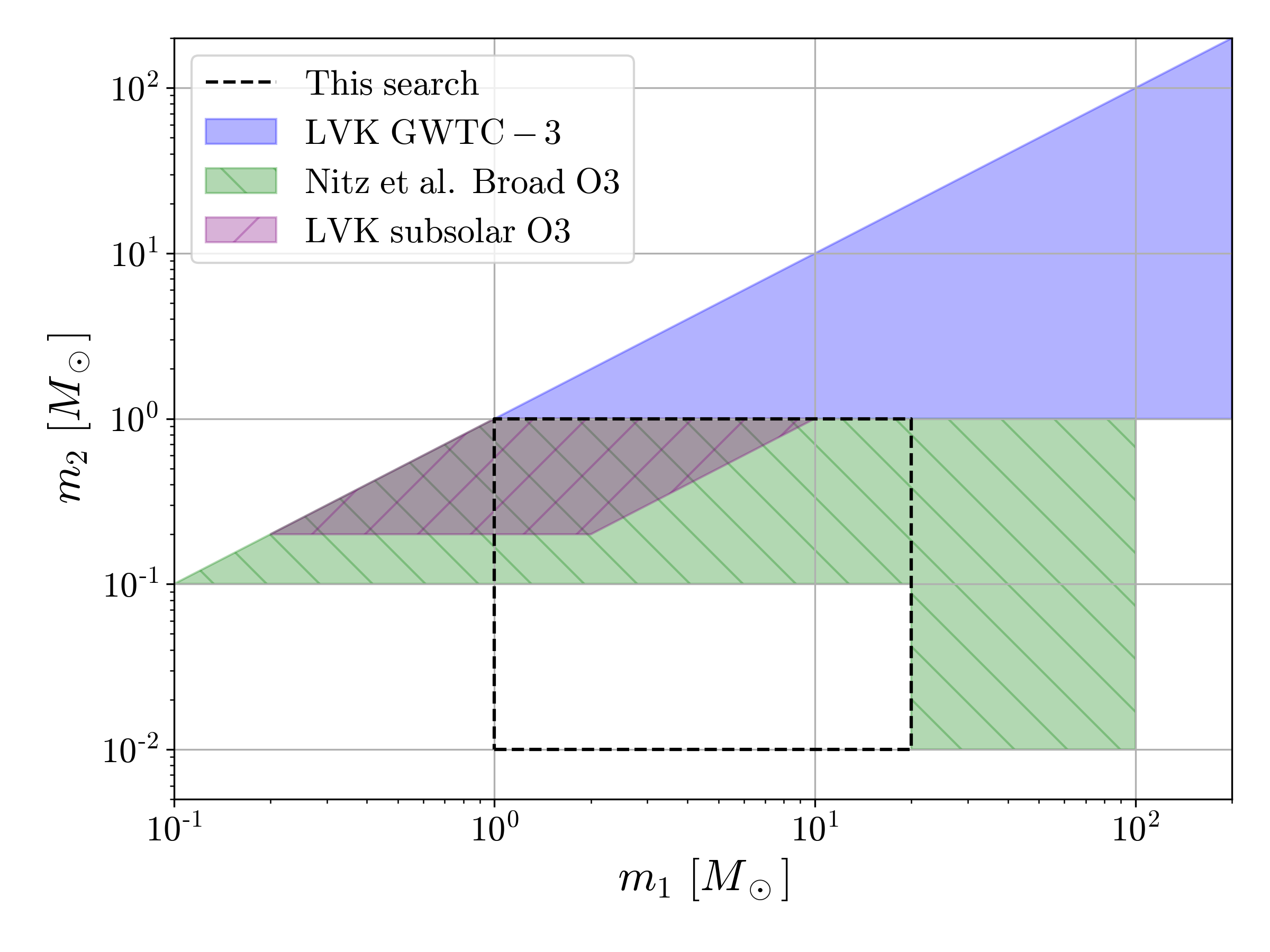}
     \caption{Region of interest compared to other recent searches. The blue region is the one searched by the LVK collaboration as part of the GWTC-3 catalog \cite{Abbott2021GWTC-3:Run}, the magenta hatched region corresponds to the LVK subsolar mass targeted search performed over O3 data \cite{Abbott2021SearchRun,O3bSubsolar} while the green hatched region corresponds to the broad search performed by Nitz and Wang over the O3 data \cite{Nitz2022Broad}.}
    \label{fig:RegionSearch}
\end{figure}

In this paper, we explore the implementation of a CNN for the identification of CBC events with very asymmetric mass configurations with $q < 2000$, and $m_1$ and $m_2$ in range between $1$ -- $20$ \msun \ and $0.01$ -- $1$ \msun, respectively.  This is motivated by the search for CBC candidates with the presence of subsolar-mass (SSM) BHs. Since there is no well-established astrophysical explanation for  the origin of SSM BHs, their discovery would point to the presence of new physics. The presence of SSM BHs are predicted by different models,  
including primordial black holes (PBHs) from the the collapse of overdensities in the early universe~\cite{Hawking:1971ei,Carr:1974nx,Hutsi2019SmallScale,Hutsi:2020sol}; gravitational collapse of dark matter halos~\cite{DAmico:2017lqj,khlopov2010primordial,belotsky2014signatures,Shandera2018DarkBlackHoles}; the accumulation of dark matter by neutron stars leading to SSM BHs~\cite{Kouvaris:2010jy}; or SSM boson stars~\cite{PhysRev.172.1331,guo2019probing,Colpi1986BosonStars}.  As illustrated in Figure~\ref{fig:RegionSearch},  this study complements the phase space in mass considered by previous searches for SSM events using O3 data and  matched-filtering based selections~\cite{Abbott2021GWTC-3:Run,Abbott2021SearchRun,Nitz2022Broad,O3bSubsolar}. Previous results using other observational periods  are included in Refs. \cite{nitz2021Ecc,nitz2021HMR,phukon2021hunt,abbott2019searchSSM}.

% =============
% DATA ANALYSIS
% =============
\section{Data Preparation}
\label{sec:da}

The study uses the O3 data from  LIGO-Hanford (H1), LIGO-Livingston (L1) and Virgo (V1) interferometers with 4096 Hz sampling rate.  
After imposing quality requirements, dealing with the understanding of the interferometer stationary noise budget as well as the identification and suppression of glitches and spectral noise contributions (for a comprehensive discussion see Refs.~\cite{davis2021ligo,acernese2022virgo}), the H1-L1-V1 combined  
samples have a total duration of 155 days. 
The H1-L1-V1 O3 data is used for constructing an image containing a spectrogram with only background and background plus injected signal for the purposes of the 
CNN training. 
Special precaution was taken in the preparation of the background images  to 
exclude any of the identified GWs events in O3, as collected in the GWTC-3 catalog \cite{Abbott2021GWTC-3:Run}. 
A  total of $142,944$ images were used. The results obtained (see below) show that this number of images is  enough for an adequate training and validation of the network. 
The images are  divided as follows: $115,200$ $(80.6\%)$ for training, $12,800$ $(9.0\%)$ for validating and $14,944$ ($10.4\%$) for testing, evenly  
distributed into background-only and background with a signal injected. 

Waveforms for GW signals are generated using PyCBC \cite{Usman2016TheCoalescence,nitz2017detecting,nitz2020gwastro} with the \textit{IMRPhenomD} \cite{Husa2016Frequency-domainSignal,Khan2016Frequency-domainEra} model and combined with data segments from the different interferometers, after taking 
into account the proper relative orientations, times of arrival and antenna factors.  The parameters considered are uniformly sampled, as described in Table~\ref{tab:TableParamSpace}, and zero spin components are assumed. Masses in the range between  1 -  20 $\msun$ (0.01 -  1 $\msun$)  are considered for $m_1$ ($m_2$), and the corresponding luminosity distance $\dlumi$ is
limited to nearby events in the range 1 - 100 Mpcs. Other parameters related to the position in the sky and orientation of the source are taken as homogeneously distributed. 

\begin{table}[htbp]
\begin{tabular}{cccccccc}
\hline \hline
$m_1~[\msun]$ & $m_2~ [\msun]$ & $\dlumi~ [$Mpc$]$ & $\psi$ & $\theta_{JN}$ & $\alpha$ & $\cos(\delta)$\\ \hline 
                $[1,20]$   &        $[0.01,1]$          &      $[1,100]$      &      $\left[0,\pi/2\right]$       &       $[0,\pi]$    & $[0,2\pi]$    & $[-1,1]$ \\ \hline \hline
\end{tabular}
\caption{Range of the uniformly sampled variables for the training set, being $(m_1,m_2)$ the component masses, $\dlumi$ the luminosity distance, $\psi$ the inclination of the orbit with respect to the line of sight, $\theta_{JN}$ the polarization of the gravitational wave and $(\alpha,\delta)$ the right ascension and declination, respectively.}
\label{tab:TableParamSpace}
\end{table}

\begin{figure}[htbp]
    \centering
    \includegraphics[width=\columnwidth]
    {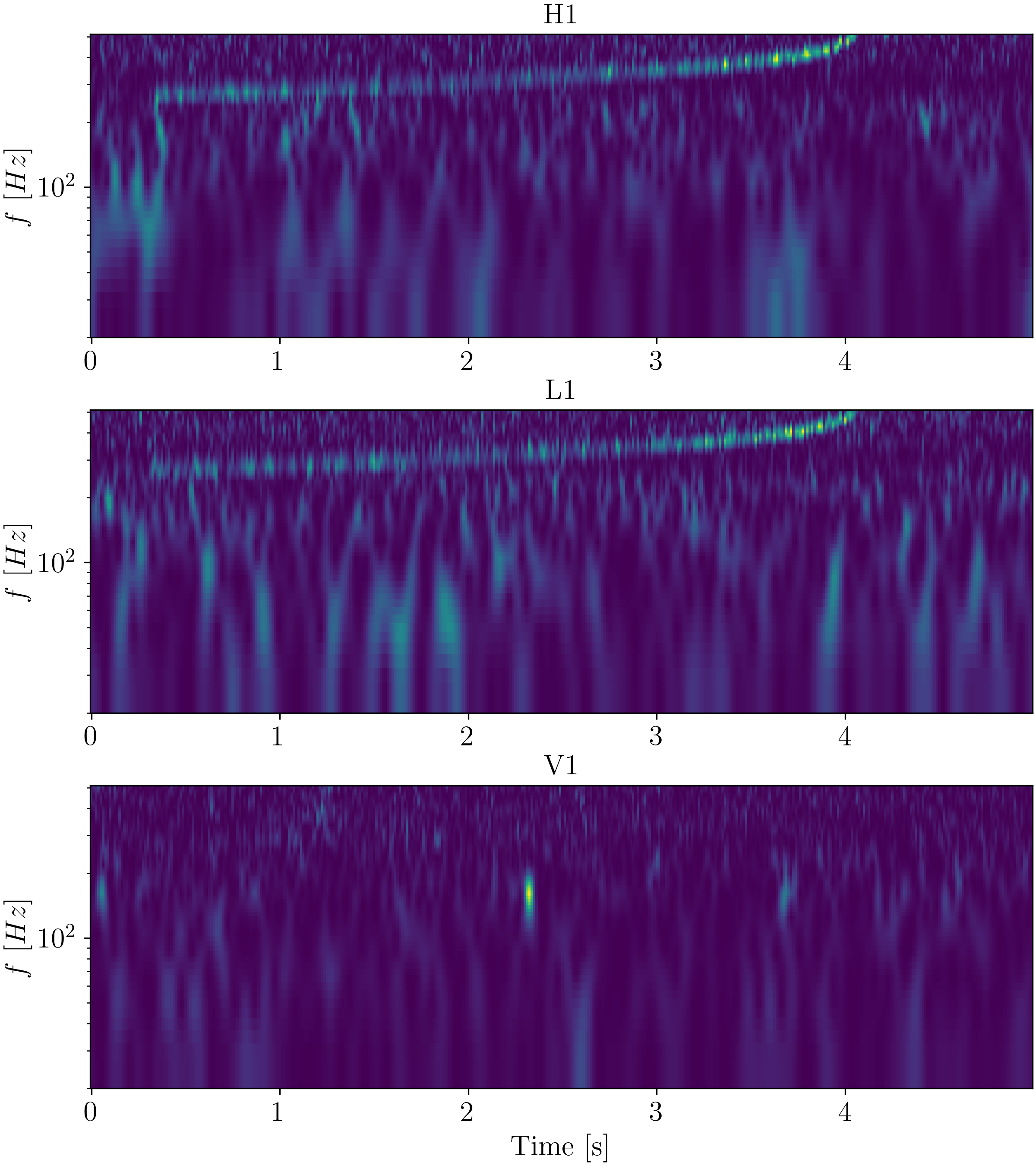}
    \caption{Spectrograms for a binary system with $m_1 = 2.6 ~\msun$ and $m_2 = 0.35~ \msun$ and a distance $\dlumi = 3.4$ Mpcs, as seen in H1, L1 and V1. }
    \label{fig:injected}
\end{figure}

The injected signals are limited to a fixed
maximum duration of  five seconds. The five-seconds window is computed backward from the merger
    time to remove low-amplitude monochromatic-like parts of the waveform
    and avoid confusing the network during training.  A low frequency threshold of 45 Hz is  applied  
 in order to control the duration of the injected signal.
Finally, the signals are randomly placed within the five-seconds window. 
Once the GW signals are injected  in the different H1, L1 and V1 background segments, the data is processed. First, the time series are whitened following the same prescription 
as in Refs.~\cite{LIGOScientific:2019hgc,DataWhitening2020}. Two-dimensional arrays holding spectrogram data are then produced using $Q$-transforms~\cite{QTransf,brown1992efficient,schorkhuber2010constant}    
in order to arrive to the desired images in terms of amplitude vs time vs frequency, with 400 bins in time and 100 bins in frequency.  Figure~\ref{fig:injected} presents an example of spectograms corresponding to a binary BH event with $m_1 = 2.6 \  \msun$ and $m_2 = 0.35 \  \msun$ at a distance of $3.4$ Mpcs. In the case of H1 and L1, the characteristic chirp is clearly observed.   

In order to account for the 
presence of glitches in the data, not completely suppressed by the whitening process and leading to instabilities in the CNN training~\cite{BatchNorm}, the contents in each image are renormalized in such a way that they have an average equal to zero and a variance equal to one, following the same prescription as in Ref.~\cite{Menendez-Vazquez2021SearchesPeriod}. 
%The whitening process followed is the one explained in Ref. \cite{DataWhitening2020} and the normalization the one presented in Ref. \cite{Menendez-Vazquez2021SearchesPeriod}.

% =============
% NN definition and training
% =============
\section{Neural network definition and training}
\label{sc:nn}

This study closely follows that of Ref.~\cite{Menendez-Vazquez2021SearchesPeriod}, using a ResNet50 architecture~\cite{DBLP:journals/corr/HeZRS15,he2016identity} (see Table~\ref{tab:arch})
with modifications in the
last layer, for which average pooling and a fully connected dense layer (1-d fc)  with a sigmoid activation function are implemented.
For the loss function, a binary cross-entropy is employed. Finally, a learning rate of $0.01$ alongside an Adam optimizer \cite{Kingma2014Adam:Optimization,ruder2016overview,goodfellow2016deep} and a batch size of $32$ are used for a total of $10$ epochs. With all these parameters, different CNNs have been trained using the GPU enhanced capabilities of \textit{Keras} and \textit{TensorFlow}~\cite{abadi2016tensorflow}. 

We train seven different CNNs.  Three CNNs are trained separately for H1, L1 and V1 data. In addition, three CNNs are trained for H1-L1, H1-V1, and L1-V1 pairs of input data, and one CNN is trained for H1-L1-V1 combined input data, where information from two or three interferometers are input simultaneously to the corresponding CNNs.
As expected, the performances of the CNNs improve by including the information of multiple interferometers during the training process, since the CNN learns about correlations across images in different channels when the signal is present. Therefore, CNNs using single interferometer information are discarded for the final scan over the O3 data. 

%
% ---- TABLE WITH CNN
%

\newcommand{\blocka}[2]{\multirow{3}{*}{\(\left[\begin{array}{c}\text{3$\times$3, #1}\\[-.1em] \text{3$\times$3, #1} \end{array}\right]\)$\times$#2}
}
\newcommand{\blockb}[3]{\multirow{3}{*}{\(\left[\begin{array}{c}\text{1$\times$1, #2}\\[-.1em] \text{3$\times$3, #2}\\[-.1em] \text{1$\times$1, #1}\end{array}\right]\)$\times$#3}
}
\renewcommand\arraystretch{1.1}
\begin{table}[htb]
\begin{center}
% \resizebox{0.7\linewidth}{!}{
\footnotesize
\begin{tabular}{c|c|c}
\hline \hline
Layer name & Output size & Layer structure  \\
\hline
conv1 & 112$\times$112 & \multicolumn{1}{c}{7$\times$7, 64, stride 2}\\
\hline
\multirow{4}{*}{conv2\_x} & \multirow{4}{*}{56$\times$56} & \multicolumn{1}{c}{3$\times$3 max pool, stride 2} \\
\cline{3-3}
  &  & \blockb{256}{64}{3} \\
  &  &  \\
  &  &  \\
\hline
\multirow{3}{*}{conv3\_x} &  \multirow{3}{*}{28$\times$28} & \blockb{512}{128}{4} \\
  &  &  \\
  &  &  \\
\hline
\multirow{3}{*}{conv4\_x} & \multirow{3}{*}{14$\times$14}  & \blockb{1024}{256}{6}\\
  &  &  \\
  &  &  \\
\hline
\multirow{3}{*}{conv5\_x} & \multirow{3}{*}{7$\times$7}  & \blockb{2048}{512}{3} \\
  &  &  \\
  &  &  \\
\hline
 & 1$\times$1  & \multicolumn{1}{c}{Global average pool, 1-d fc, sigmoid} \\
\hline \hline
%\multicolumn{2}{l|}{FLOPs} & 3.8$\times10^9$ \\
%\hline
\multicolumn{3}{l}{Hyper parameters} \\ \hline
\multicolumn{2}{l}{Learning rate} & 0.01 \\
\multicolumn{2}{l}{Batch size} & 32 \\
\multicolumn{2}{l}{Number of epochs} & 10 \\
\multicolumn{2}{l}{Optimizer} & Adam \\
\multicolumn{2}{l}{Loss function} & Binary-cross entropy \\ \hline \hline
\end{tabular}
% }                                                                                                                                                                                                                                        
\end{center}
\vspace{-.5em}
\caption{
CNN architecture and the associated hyper parameters. Building blocks are shown in brackets, with the numbers of blocks stacked.
Downsampling is performed by conv3\_1, conv4\_1, and conv5\_1 with a stride of 2 (partially taken from~\cite{DBLP:journals/corr/HeZRS15}).
}
\label{tab:arch}
\vspace{-.5em}
\end{table}

\begin{figure}[htbp]
    \centering
    \includegraphics[width=\columnwidth]
    {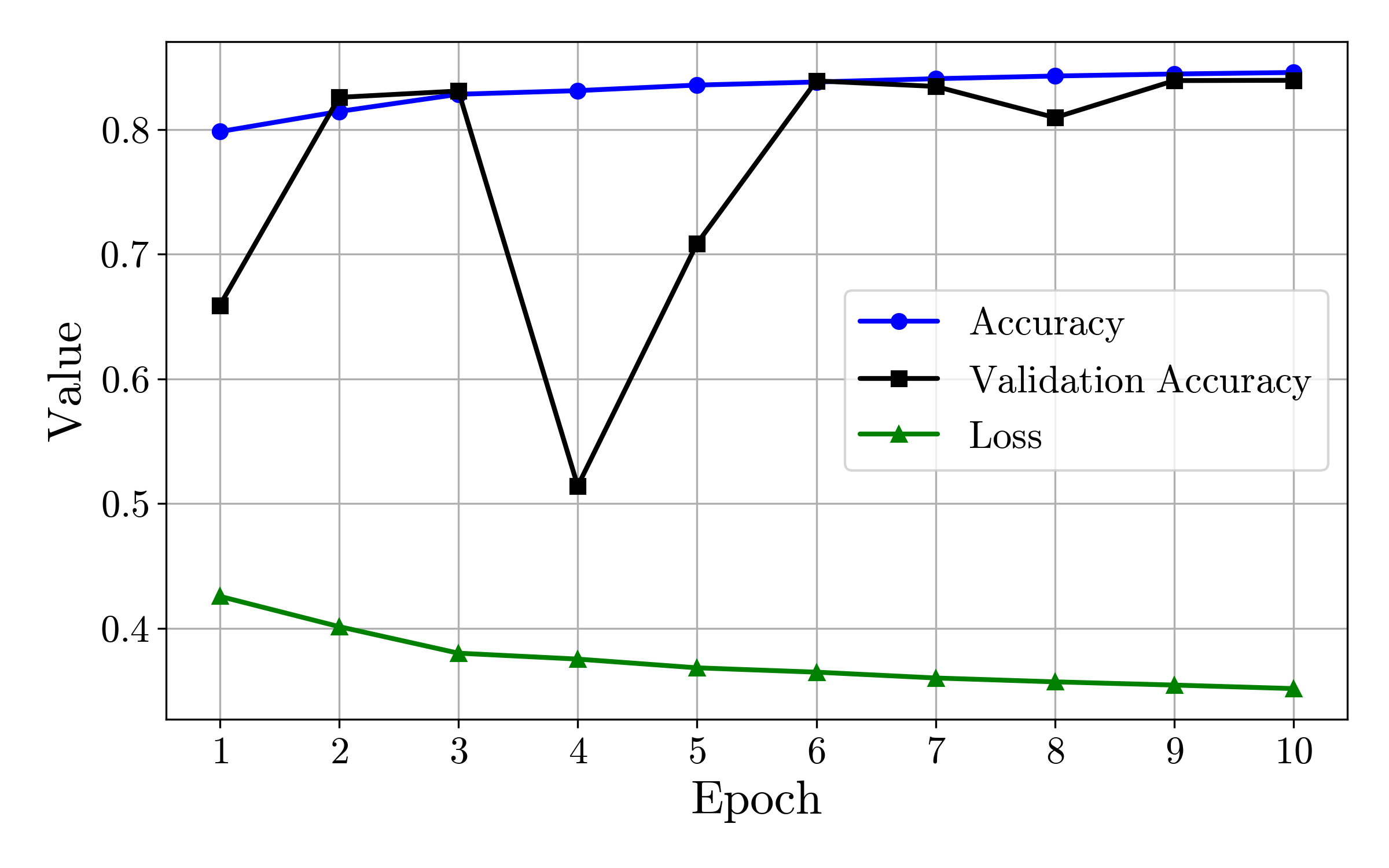}
    \caption{Accuracy, loss and validation accuracy during the training epochs for the H1-L1-V1 CNN.}
    \label{fig:CNNtrain}
\end{figure}

\begin{figure}[htbp]
    \centering
    \includegraphics[width=\columnwidth]
    {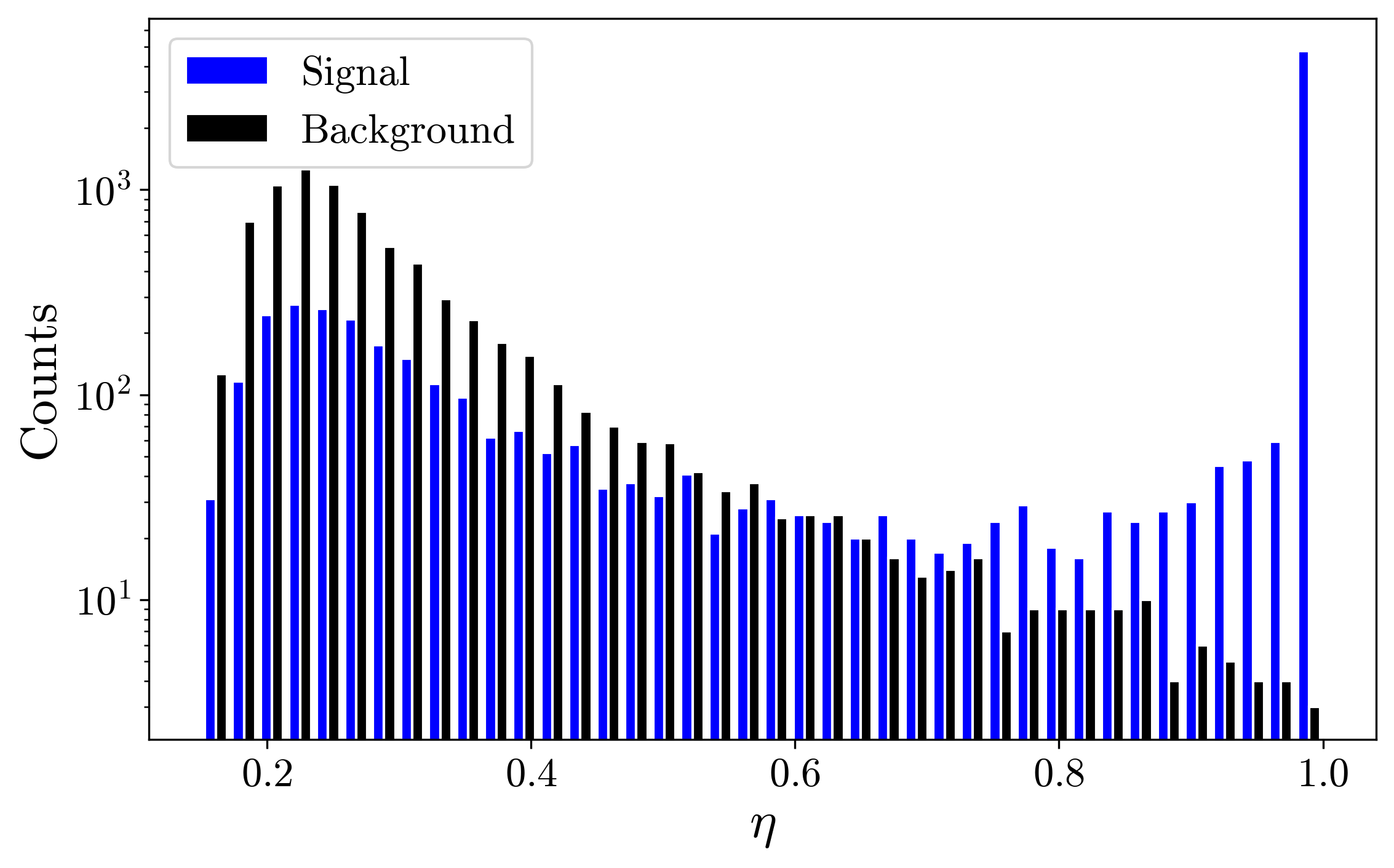}
    \caption{CNN discriminating output corresponding to the H1-L1-V1 case for background and signal images.}
    \label{fig:CNNout}
\end{figure}

Figure~\ref{fig:CNNtrain} shows, for the H1-L1-V1 case,  the evolution of the accuracy and loss as a function of epochs, demonstrating stability after about eight to ten epochs,  with an accuracy above 0.8 and a loss below 0.4.  In addition, the validation accuracy is presented, demonstrating a healthy evolution of the training process.  The final CNN output for the H1-L1-V1 case is shown in Figure~\ref{fig:CNNout}, where a clear discrimination is obtained between signal and background samples. Similar features in the training process and the distribution of the final CNN discriminant are observed in the rest of CNNs.

\begin{figure}[htbp]
    \centering
    \includegraphics[width=\columnwidth]
    {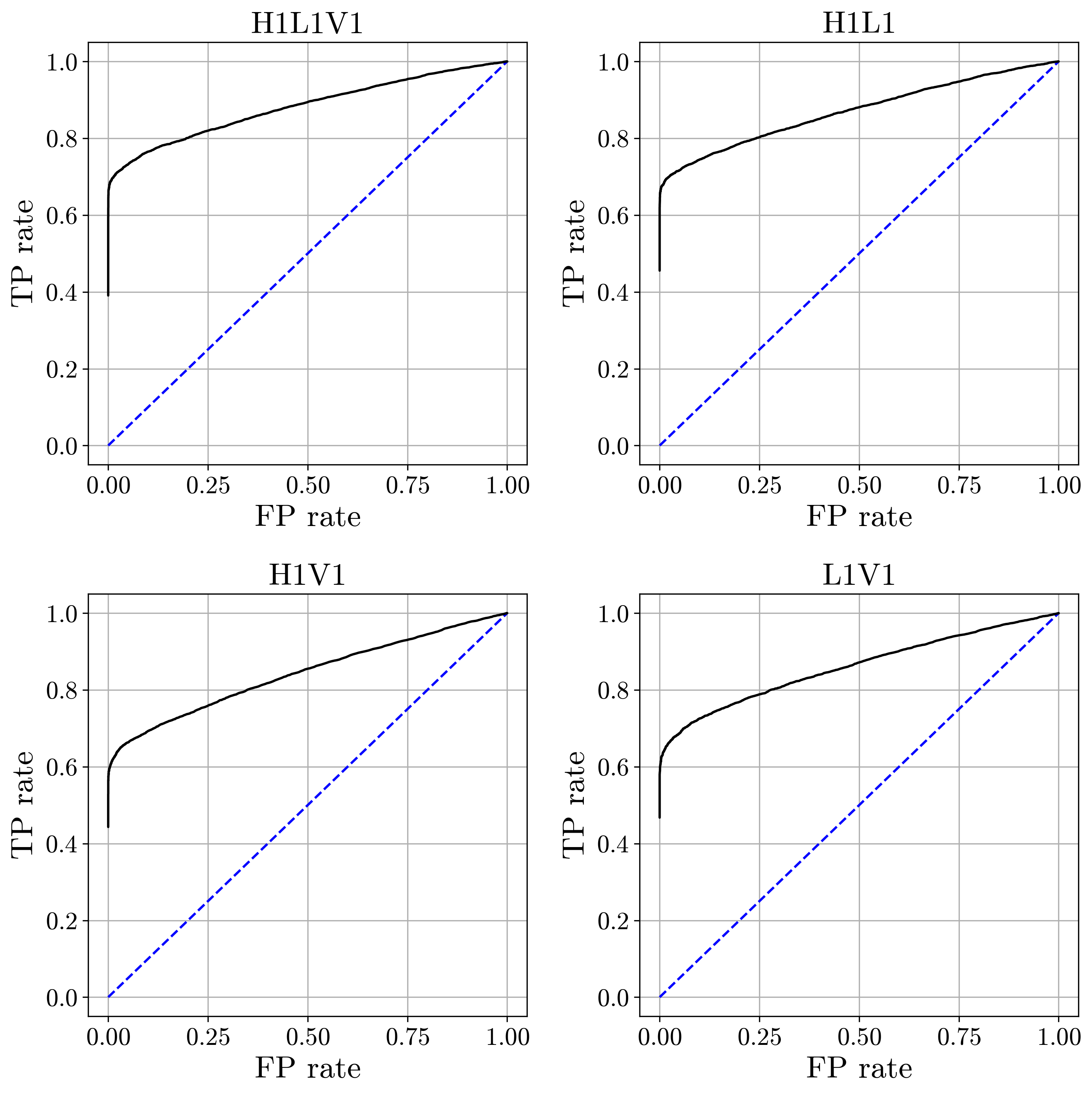}
    \caption{ROC curves  for the different CNNs.}
    \label{fig:roc}
\end{figure}

The receiver operating characteristic (ROC)  curves for the separate CNNs,   
representing the true positive (TP) versus the false positive (FP) rates are presented in  
Figure~\ref{fig:roc} for the H1-L1, H1-V1, L1-V1, and H1-L1-V1 CNNs.  For very low FP rates, the TP rates only reach values around $70\%$, indicating a limited efficiency for event detection.  The efficiency steadily increases at the cost of much larger FP rates.   
The ROC curve, along with a tolerable maximum false alarms rate (FAR) for detection, determines the final operating point %$\eta_0$
of a given CNN. The computation of the FAR for each CNN follows the prescription in Ref.~\cite{Abbott2016ObservationMerger}. The FAR is defined as ${\rm{FAR (\eta)}} = N(\eta)/T$, where $\eta\in[0,1]$ is the CNN discriminant output, $N(\eta)$ is the number of events with a CNN discriminant above or equal to $\eta$ and $T$ the period of time analysed.  In order to effectively increase the time considered in the calculation, reaching very low FAR values, the time slide technique \cite{Abbott2005SearchStars,Abbott2016ObservationMerger} is used. This allows accumulating $\mathcal{O}(10^9)$ images of $5$ s duration each and accessing FAR values down to $1/152.6$~years${}^{-1}$. 

\begin{figure}[htbp]
    \centering
    \includegraphics[width=\columnwidth]
    {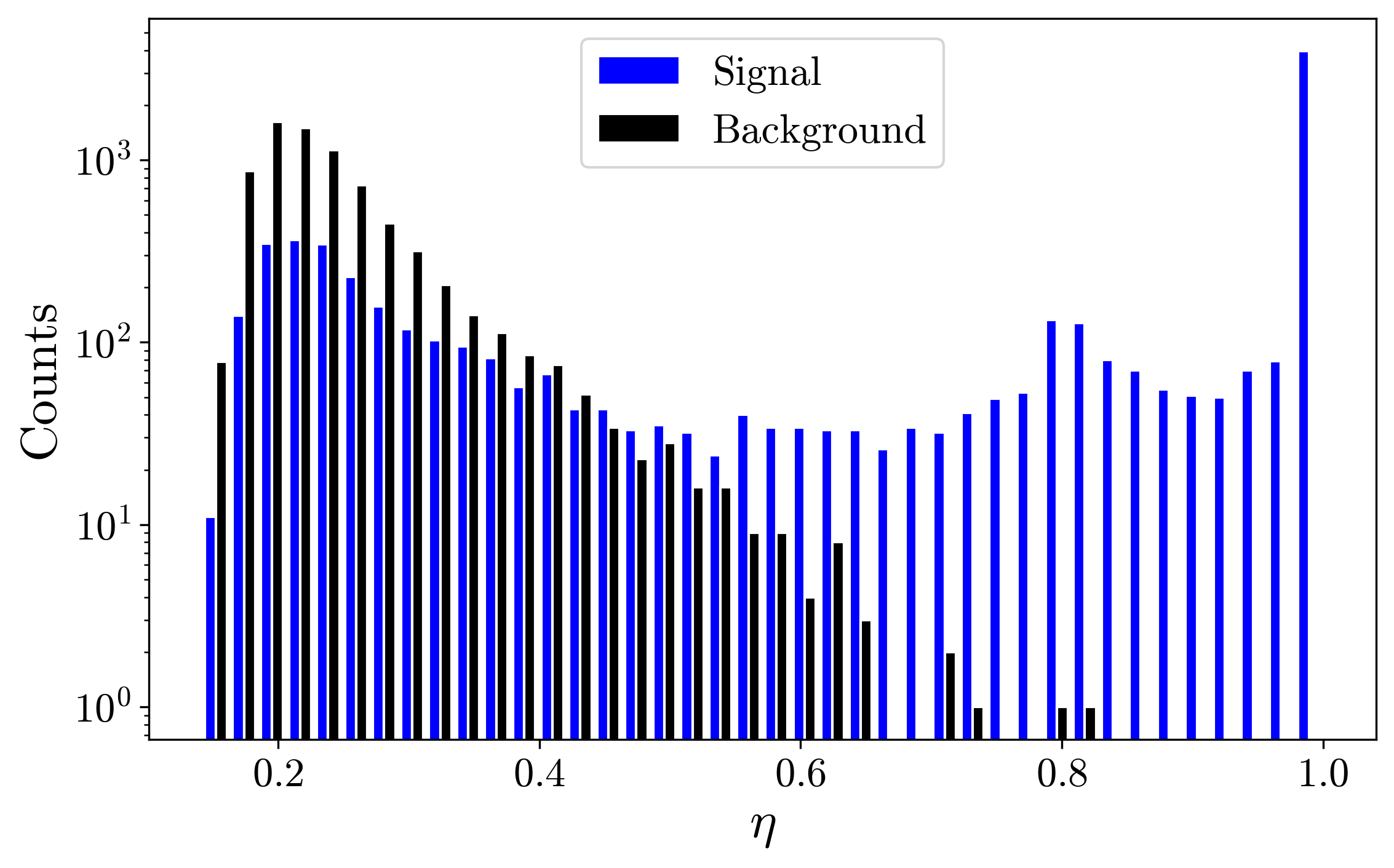}
    \caption{CNN global discriminating output corresponding to the average of the H1-L1-V1,  H1-L1, L1-V1, and H1-V1 CNN outputs for background and signal images.}
    \label{fig:CNNcomb}
\end{figure}

We initially establish a CNN discriminant $\eta_0$ corresponding to a $\rm{FAR (\eta_0)}$ value of $1$~years${}^{-1}$.
% Marc addition:
However, the number of FP detected remains sizable when $\eta_0 \to 1$ and the CNNs  never reach a discriminant capable of producing only one false positive event per year. A further improvement of the global sensitivity is
achieved by combining the outputs of the separate CNNs into a global 
discriminant. Such combination provides an additional tool for suppressing glitches in the data affecting independently
the interferometers and in different time stamps.  A simple average of the H1-L1-V1, H1-L1, L1-V1, and H1-V1 CNN outputs has been considered. Alternatively, a number of algorithms, potentially giving different weights to different CNNs, were explored leading to very similar or even worse results.
The resulting discriminant is presented in Figure~\ref{fig:CNNcomb} demonstrating an improved separation between background and signal, leading to a higher  significance for the events finally selected as signal.  Table~\ref{tab:nnth} collects the corresponding detection rates and the computed FAR upper limit in the case of $\eta_0 = 1$ for the separate CNNs and their combination, where only the latter shows FAR values less than 1 event per year. 
   
%Since the number of false positives detected during the FAR estimation procedure, these CNNs never reach a discriminant capable of producing only one false positive event per year and the limit FAR is therefore included in Table~\ref{tab:nnth}. This implies that when the discriminant equals $1.0$ the corresponding significance would be $FAR\leq FAR_l$, without the possibility of narrowing it down.
 
\begin{table}[htb]
\begin{center}
\begin{footnotesize}
\begin{tabular}{l c c c c} \hline \hline
% \multicolumn{4}{c}{single interferometer channel} \\ \hline
CNN & Threshold & TP rate & FP rate & $\rm{FAR (\eta_0 = 1)}$ $[$yrs$^{-1}]$\\ \hline
% L1 & & & \\
% H1 & & & \\
% V1 & & & \\ \hline
%
% -----
%
%\multicolumn{4}{c}{double interferometer channel} \\ \hline
% & CNN discriminant ($\%$) & TP rate ($\%$) & FP rate ($\%$) \\ \hline
H1 -- L1 & $1.0$ & $0.46$ & $\leq 2\cdot 10^{-4}$ & $\sim 10^2$\\%$175$\\
L1 -- V1 & $1.0$ & $0.47$ & $\leq 2\cdot 10^{-4}$ & $\sim 10^3$\\%2238\\
H1 -- V1 & $1.0$ & $0.44$ & $\leq 2\cdot 10^{-4}$ & $\sim 10^2$\\ %81\\
\hline
%
% -----
%
%\multicolumn{4}{c}{triple interferometer channel} \\ \hline
% & CNN discriminant ($\%$) & TP rate ($\%$) & FP rate ($\%$) \\ 
H1 -- L1 -- V1 & $1.0$ & $0.58$ & $\leq 2\cdot 10^{-4}$ & $\sim 10^{3}$\\ %1063\\
 \hline
 
Combined & $0.998$ & $0.50$ & $\leq 2\cdot 10^{-4}$ & $  \sim 10^{-2}$\\ % \leq 0.01
\hline \hline
\end{tabular}
\end{footnotesize}
\caption{\small 
Anticipated TP and FP rates and FAR for the different CNNs and a discriminant $\eta_0 = 1$. The FP rates represent a 95\% confidence level upper limits according to a null observation of FP in the testing set assuming Poisson statistics.
}
\label{tab:nnth}
\end{center}
\end{table}

\begin{figure}[htbp]
    \centering
    \includegraphics[width=\columnwidth]
    {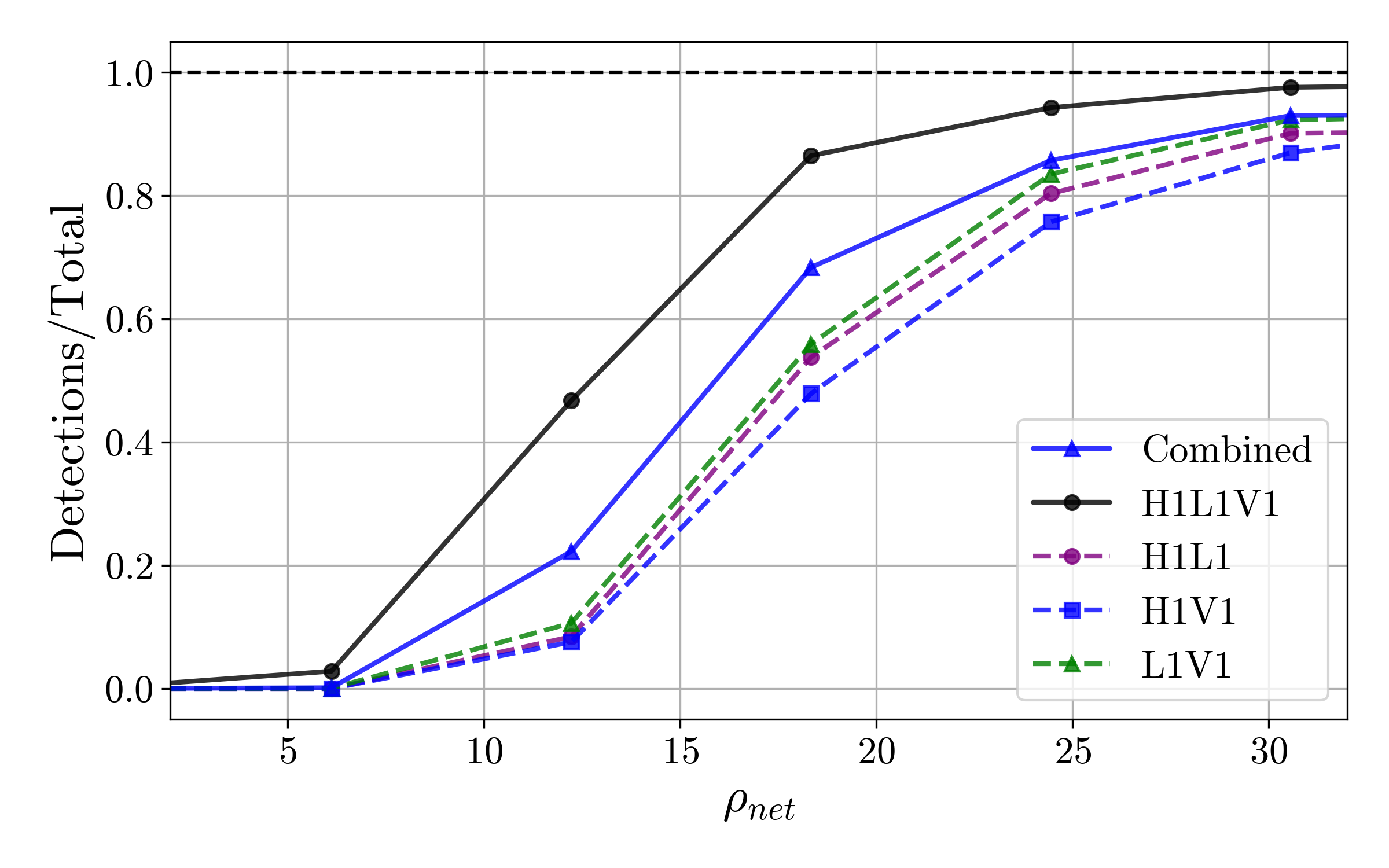}
    \caption{Signal detection efficiency as a function of  $\rho_{\rm net}$ for the different CNNs and their combination.}
    \label{fig:SNRnet}
\end{figure}

% ===============
% INJECTION TESTS
% ===============
Signal injection studies are performed to establish the sensitivity of the different CNNs to the presence of a GW signal. For each GW signal, the signal-to-noise ratio (SNR), $\rho$, is computed following the 
prescription in Ref.~\cite{PhysRevLett.120.141103} solving the integral 
\begin{equation}
    \rho^2 = 4 \   \int_{f_{\rm{min}}}^{f_{\rm{max}}}  \td f \ \frac{|\tilde{h}(f)|^2}{S_n(f)}, 
\end{equation}
\noindent
in the frequency domain $(f)$, where $\tilde{h}(f)$ denotes the signal in the frequency domain and $S_n(f)$ 
 the power spectral density of the background. A Tukey window with $\alpha = 1/9$ is considered 
for the Fourier transform. The SNR defined above refers to each of the interferometers separately.  Following the work in Refs.~\cite{Littenberg2016EnablingBursts,Lee2021EnhancingPipeline}, when appropriate we define a network SNR, $\rho_{\rm net}$, as
\begin{equation}
    \rho_{\rm net}^2 =  \sum_i \rho_i^2 \ , 
\end{equation}
\noindent
where $i$ denotes the different interferometers. Figure~\ref{fig:SNRnet} shows the fraction of GW signals identified by the CNNs as a function of $\rho_{\rm net}$ in the different cases.   As expected, the efficiency for signal detection increases rapidly with increasing SNR,  becoming more efficient 
for large $\rho_{\rm net}$ values and improving with the inclusion of information from multiple interferometers.   
The best results are obtained by the H1-L1-V1 CNN. The results from the combination of CNNs is a compromise between the H1-L1-V1 CNN and the rest.  Events with $\rho_{\rm net} \geq 25$ would be detected with an efficiency above 95$\%$ in the case of the H1-L1-V1 CNN. Table~\ref{tab:rhonet} collects the values of  $\rho_{\rm net}$ at given detection efficiencies for the different CNNs.

\begin{table}[htb]
\begin{center}
\begin{footnotesize}
\begin{tabular}{l c c c} \hline \hline
CNN & $\rho_{\rm net} (80\%)$ & $\rho_{\rm net} (90\%)$ & $\rho_{\rm net} (95\%)$\\ \hline
% L1 & & & \\
% H1 & & & \\
% V1 & & & \\ \hline
%
% -----
%
%\multicolumn{4}{c}{double interferometer channel} \\ \hline
% & CNN discriminant ($\%$) & TP rate ($\%$) & FP rate ($\%$) \\ \hline
H1 -- L1 & $24.4$ & $30.5$ & $42.0$\\
L1 -- V1 & $23.7$ & $29.0$ & $41.2$\\
H1 -- V1 & $26.8$ & $21.1$ & $25.8$\\ \hline
%
% -----
%
%\multicolumn{4}{c}{triple interferometer channel} \\ \hline
% & CNN discriminant ($\%$) & TP rate ($\%$) & FP rate ($\%$) \\ 
 H1 -- L1 -- V1 & $17.3$ & $21.1$ & $25.8$\\ \hline
  Combined & $22.4$ & $28.0$ & $40.1$\\ \hline \hline
\end{tabular}
\end{footnotesize}
\caption{\small 
Values of $\rho_{\rm net}$ at given detection efficiencies for the different 
CNNs and the combination of outputs.
}
\label{tab:rhonet}
\end{center}
\end{table}

% ==============
% RESULTS
% ==============
\section{Results}
We performed a  scan of the full O3 data set, using the H1-L1-V1 combined sample, 
for which a slicing window of five seconds duration was used in steps of 2.5 seconds 
(leading to a 50$\%$ overlap between consecutive images) in each of the interferometers. 
This translates into more than eighty  million images to be tested for the presence of potential signals.
The CNN global discriminating output, defined as the average of the H1-L1-V1,  H1-L1, L1-V1, and H1-V1 CNN outputs, is used to search for signal of SSM events.  A scan over the data using different global discriminating values in the range between 0 and 1 is performed.  In each case, the corresponding FAR is computed.  
The computation time for the entire O3 scan has been of the order of $2,000$ CPU-hours (on an Intel$^\text{\textregistered}$ Xeon$^\text{\textregistered}$ CPU E5-2680 v4 @ 2.40GHz). 
This represents a major improvement compared with the typically required CPU-time for 
known matched filtering pipelines.

\begin{figure}[htbp]
    \centering
    \includegraphics[width=\columnwidth]{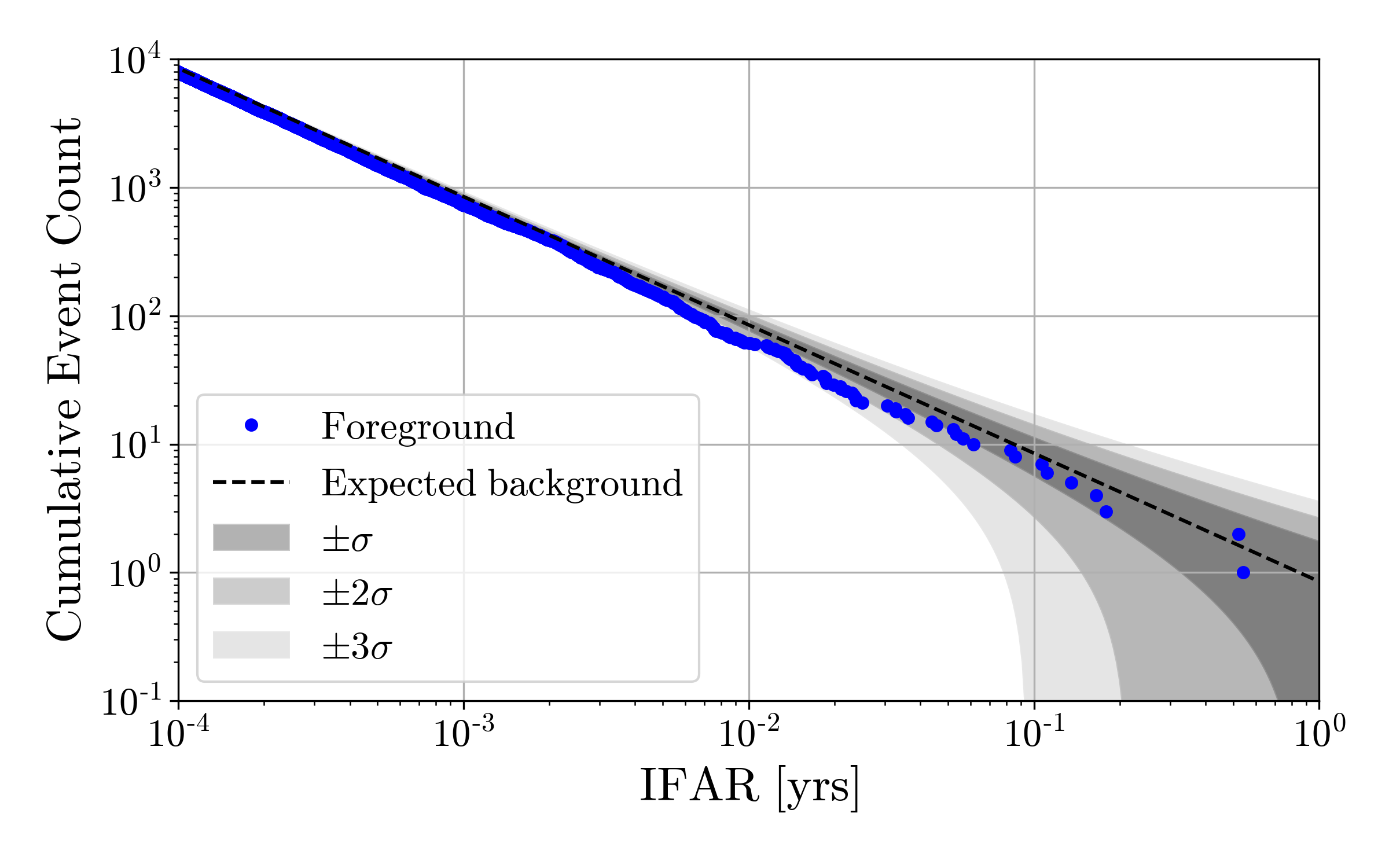}
    \caption{Observed cumulative event count as a function of the inverse false alarm rate for the O3 scan (dots). The data is compared to foreground predictions (dashed line) and including 1$\sigma$, 2$\sigma$, and 3$\sigma$ bands (shadowed areas). }
    \label{fig:ECvsIFAR}
\end{figure}

\begin{figure}[htbp]
    \centering
    \includegraphics[width=\columnwidth]{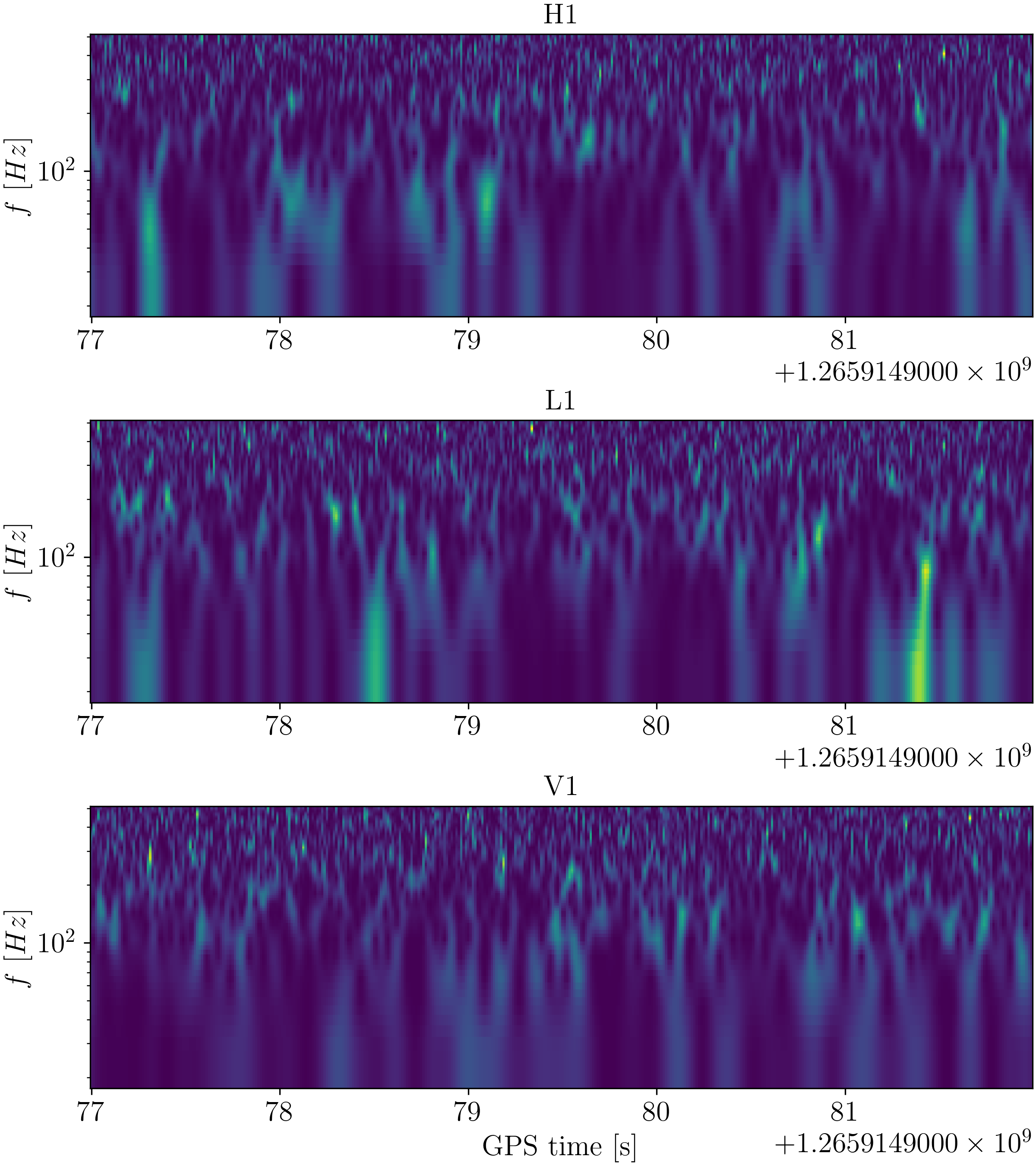}
    \caption{Spectrogram of the most significant image. Corresponds to a $\mathrm{FAR}=1.9~\mathrm{yrs}^{-1}$}
    \label{fig:MostSignificantEvent}
\end{figure}

The resulting  inverse FAR distribution  (IFAR), in units of years,  is presented in Figure~\ref{fig:ECvsIFAR} compared to the expected yield of noise events following a Poisson probability distribution. 
No  significant deviation from the expected noise is observed and no claim of SSM event detection can be made.  For illustration purposes, Figure~\ref{fig:MostSignificantEvent} shows the H1, L1 and V1 spectrograms for the most significant event having a FAR of $1.9$~years${}^{-1}$, a combined CNN value equal to 0.9635, and CNN values equal to 0.9848, 0.9172, 0.9774 and 0.9747  for the H1-L1-V1, H1-L1, L1-V1, and H1-V1 neural networks, respectively.

\begin{figure}[htbp]
    \centering
    \includegraphics[width=\columnwidth]{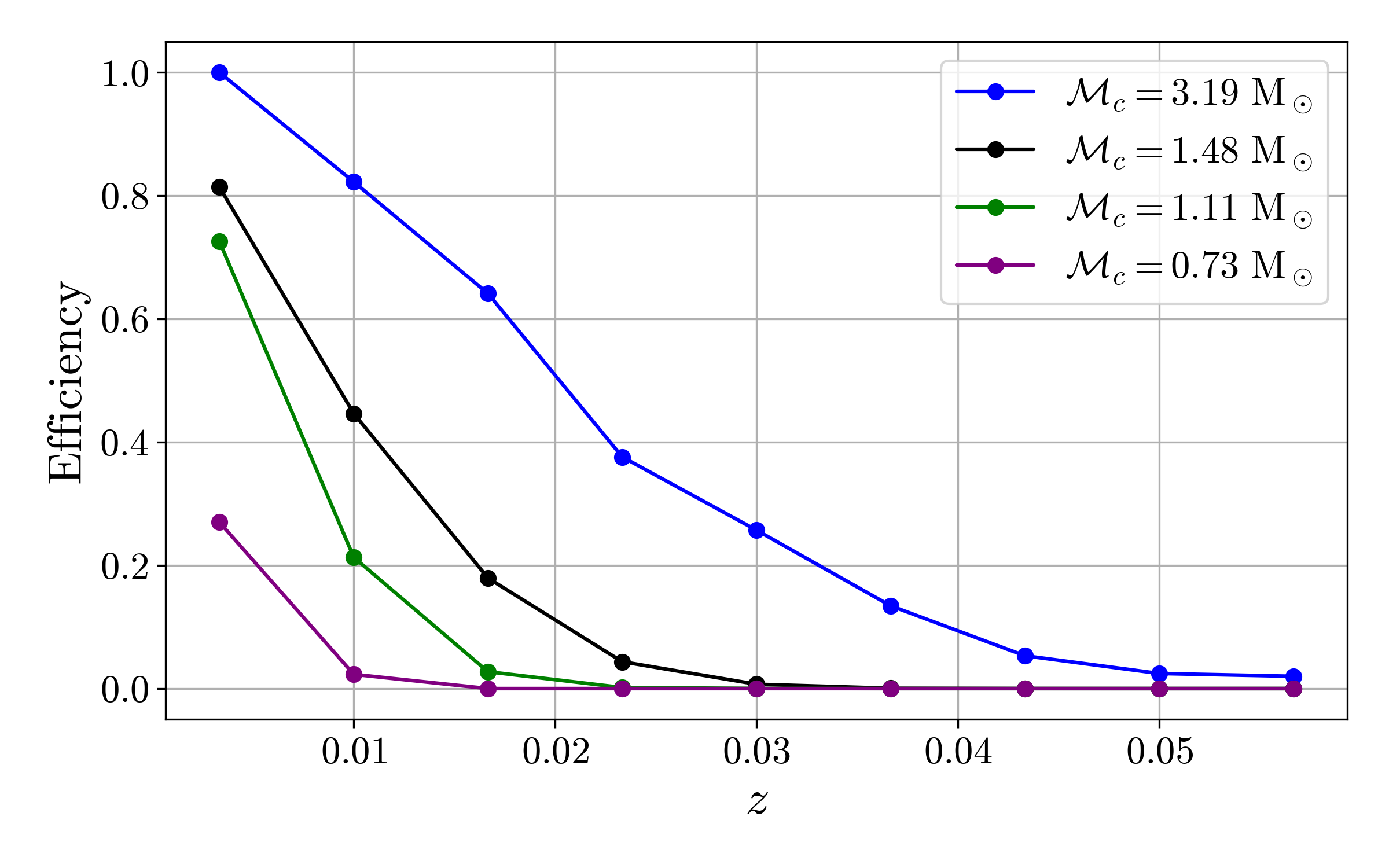}
    \caption{Detection efficiency for the combined CNN discriminant as a function of $z$ for different values of the chirp mass.}
    \label{fig:eff}
\end{figure}

The results are translated into 90$\%$ confidence level (CL) upper limits of the merger rate of binary systems in the range of masses and $q$ values considered.  Since the sensitivity for detection mostly depends on the chirp mass of the binary system, defined as $\mathcal{M}_c \equiv \frac{(m_1 m_2)^{3/5}}{(m_1 + m_2)^{1/5}}$, the computed merger rates are binned in $\mathcal{M}_c$ instead of in the single masses of the binary system.  The 90$\%$ CL upper limits are calculated using the loudest event statistics approach~\cite{tiwari2018estimation,Abbott2021SearchRun,Nitz2021SearchRun,Nitz2022Broad,O3bSubsolar} in terms of the surveyed time-volume $\langle VT \rangle$, following the expression

\begin{equation}
    \mathcal{R}_{\rm{90}} = \frac{2.3}{\langle VT \rangle }, \ \langle VT \rangle = T \ \int \td z \frac{1}{1+z} \frac{\td V_c}{\td z} \varepsilon(z), 
\end{equation}
\noindent
where $T$ is the total observation time, $z$ denotes the redshift, $V_c$ is the comoving volume and $\varepsilon$ is the efficiency for detection.  In this study $T$ is limited to $155$ days when H1, L1 and V1 interferometers were all taking data simultaneously. 
Figure~\ref{fig:eff} presents the detection efficiency of the combined CNN discriminant as a function of $z$ in several $\mathcal{M}_c$ 
bins. The efficiency is computed using injected signals and it vanishes for $z > 0.06$. The integral above is marginalized over the rest of parameters of the binary system (see Table~\ref{tab:TableParamSpace}),  which are considered homogeneously distributed in comoving volume.  

\begin{figure}[htbp]
    \centering
    \includegraphics[width=\columnwidth]{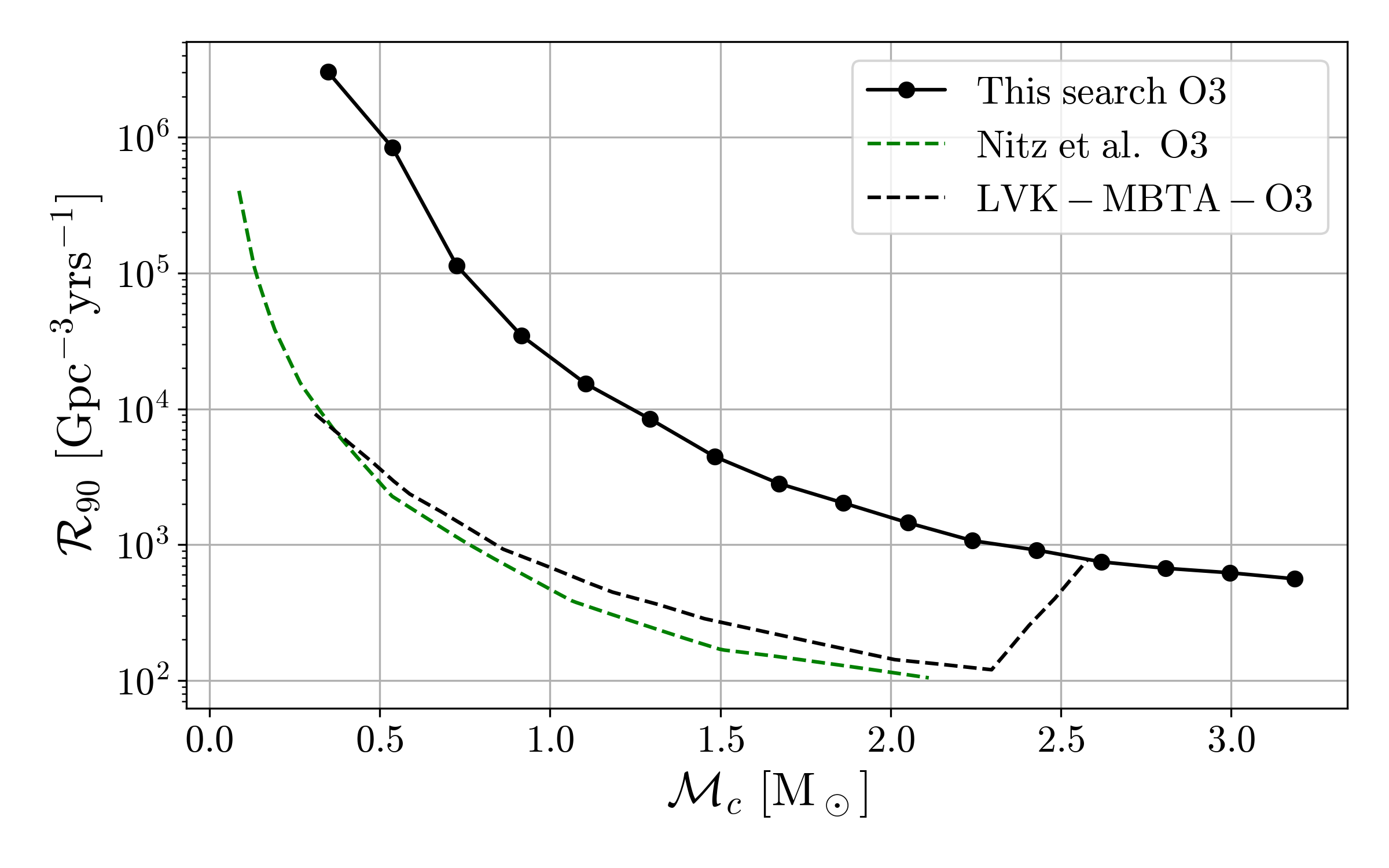}
    \caption{The 90$\%$ confidence level upper limit on $\rr$ as a function  of the chirp mass. Our result (solid line) is compared with matched-filtering based results~\cite{O3bSubsolar,Nitz2022Broad} (dashed lines).}
    \label{fig:R90}
\end{figure}

\begin{figure}[htbp]
    \centering
    \includegraphics[width=\columnwidth]{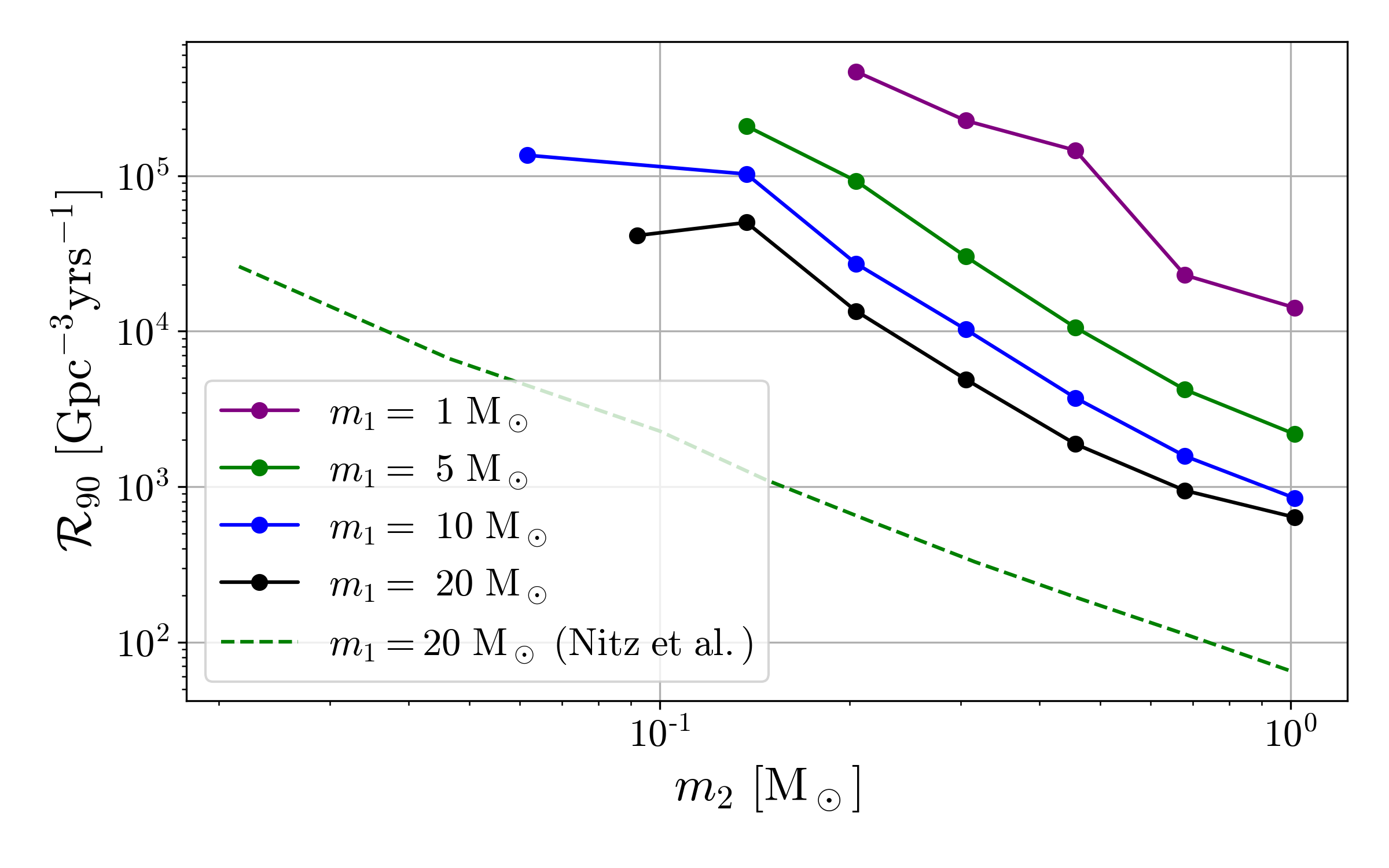}
    \caption{The 90$\%$ confidence level upper limit on $\rr$ as function  of $m_2$ in  different $m_1$ regions. Our result (solid lines) is compared with Nitz et al. results~\cite{Nitz2022Broad} (dashed line).}
    \label{fig:R90m1m2}
\end{figure}

Figures~\ref{fig:R90} and  ~\ref{fig:R90m1m2} presents the 90$\%$ CL upper limits on the merging rate as a function of the chirp mass and as a function of  $m_2$ in different $m_1$ regions, respectively.   The results are compared to similar results from matched-filtering based analyses.  Our result provides 90$\%$ CL upper limits in the range between $3\times 10^6~\mathrm{Gpc}^{-3}\mathrm{yrs}^{-1}$ and $560~\mathrm{Gpc}^{-3}\mathrm{yrs}^{-1}$ with increasing chirp mass, extending previous results to chirp masses up to 3 $\msun$.  At lower chirp mass, our  constrains are weaker than previous results.  This is partially attributed to the effective reduction of the observation time,  by a factor of about two,  from limiting the data to simultaneous H1-L1-V1 configurations, as a way to obtain manageable false alarm rates. As shown in Figure~\ref{fig:R90m1m2}, the constrains from our analysis are more stringent with increasing mass difference $m_1 - m_2$, as expected for a CNN trained on very asymmetric configurations.  

% ==============
% SUMMARY
% ==============
\section{Conclusions}
We present the results on a search for compact binary coalescence events with asymmetric mass configurations with masses in the range $0.01~\msun$  and $1~\msun$ for the lighter object and between $1~\msun$  and $20~\msun$ for the heavier, using LIGO-Virgo O3 data and dedicated convoluted neural networks based on the analysis of frequency-time spectrograms. Different neural networks and combination of them are explored,  involving the simultaneous use of several interferometers data. The scan over the O3 data results into no significant signal events found.  
The CNN approach for scanning the data is found to be much faster than traditional matched-filtering based pipelines.  
The CNN results  are translated into 90$\%$ confidence level upper limits on the merger rates as a function of the mass parameters of the binary system for events within $z < 0.06$ and for the trained range.  Although the results do not improve other bounds using matched-filtering techniques, partially due to the limited observation time considered, the CNN approach allows for effectively extending the search towards larger chirp masses.
%Marc: New paragraph
%This approach is also faster than traditional pipelines and can be deployed in either CPU or GPU machines.

% ======================
% Acknowledgments
% ======================
\section*{Acknowledgements}

This material is based upon work supported by NSF’s LIGO Laboratory which is a major facility
fully funded by the National Science Foundation.
The authors also gratefully acknowledge the support of
the Science and Technology Facilities Council (STFC) of the
United Kingdom, the Max-Planck-Society (MPS), and the State of
Niedersachsen/Germany for support of the construction of Advanced LIGO 
and construction and operation of the GEO\,600 detector. 
Additional support for Advanced LIGO was provided by the Australian Research Council.
The authors gratefully acknowledge the Italian Istituto Nazionale di Fisica Nucleare (INFN),  
the French Centre National de la Recherche Scientifique (CNRS) and
the Netherlands Organization for Scientific Research (NWO), 
for the construction and operation of the Virgo detector
and the creation and support  of the EGO consortium. 
The authors thankfully acknowledge the computer resources at MinoTauro and the technical support provided by Barcelona Supercomputing Center (RES-FI-2021-3-0020). MAC is supported by the 2022 FI-00335 grant. This paper has been given LIGO DCC number LIGO-P2200184-v2. This work is partially  supported   by  the Spanish MCIN/AEI/ 10.13039/501100011033  under
the grants SEV-2016-0588, PGC2018-101858-B-I00,  and PID2020-113701GB-I00 some of which include ERDF  funds  from  the  European  Union. IFAE  is  partially funded by the CERCA program of the Generalitat de Catalunya. This work was carried out within the framework of the EU COST action No. CA17137.

% ==============
% BIBILIORGAPHY
% ==============

\bibliographystyle{apsrev}
\bibliography{ref}{}

% ===============
% PAPER ENDS
% ===============

\end{document}